\documentclass[a4paper,11pt]{article}
\usepackage{jheppub} % for details on the use of the package, please see the JINST-author-manual
\usepackage{lineno}
\usepackage{subcaption}
\usepackage{booktabs}

\title{\boldmath Constraining new physics in charm quark associated Higgs boson production events using the Standard Model effective field theory approach}

\author[1]{Nordin Breugelmans\note{Corresponding author.},}
\author{Felix Heyen,}
\author[2]{Jorgen D'Hondt\note{Now at Nikhef, Science Park 105, 1098 XG Amsterdam, The Netherlands.},} 
\author{Michael Tytgat,}
\author{Gerrit Van Onsem}
\affiliation{Inter-University Institute for High Energies, Vrije Universiteit Brussel, Pleinlaan 2, 1050 Brussels, Belgium}

\emailAdd{Nordin.Breugelmans@vub.be}

\abstract{As the search for observable deviations from the Standard Model of particle physics remains to be of significant interest, effective field theory (EFT) continues to be a popular method to parametrize such effects. In this work, a first-time investigation is performed of the unique capability of measurements of charm quark associated Higgs boson production (cH) in proton-proton collisions at the CERN Large Hadron Collider to constrain a set of dimension-six EFT operators. The phenomenology of these operators is discussed and a proposed analysis strategy is presented, with a focus on $\mathrm{H}\rightarrow \mathrm{Z}\mathrm{Z}^{*}\rightarrow 4\mu$ decays, using a generic detector simulation that is parametrized to reflect the response of the CMS detector at the LHC. From this, expected 95\% CL upper limits are derived for the Wilson coefficients of individual operators by considering yield and shape effects in the spectra of the four-muon invariant mass $m_{4\mu}$ and leading jet transverse momentum $p_{T}$. Scenarios with simultaneous contributions from two operators are also considered. Finally, potential analysis improvements that may be implemented in an experimental context are outlined.}

\begin{document}
\maketitle
\flushbottom

\section{Introduction}
\label{sec:intro}
The discovery of the Higgs boson by the CMS \cite{CMSHiggsDiscovery1,CMSHiggsDiscovery2} and ATLAS \cite{ATLASHiggsDiscovery} experiments at the Large Hadron Collider (LHC) at CERN is commonly regarded as a key milestone within the Standard Model (SM) of particle physics. This experimental result marked the beginning of a new era of probing the scalar sector of the SM to further our understanding of the mechanisms through which particle masses are generated. To this end, many of the properties of the Higgs boson, such as the mass, spin, width, and production and decay rates, have been measured by the ATLAS and CMS experiments \cite{ATLASCMSHiggsMeasurements,ATLASTenYearsHiggs,CMSTenYearsHiggs}. Furthermore, the couplings of the Higgs boson to the heavy vector bosons, top quark, bottom quark, tau lepton, and muon have been found to be consistent with the SM prediction \cite{CombinedHiggsCouplings,CMS-Higgs-couplings-muons}. In recent years, efforts have been made to also constrain the Higgs-charm Yukawa coupling at the LHC by probing the direct decay $H\rightarrow c\bar{c}$ in the $VH$ and $ttH$ production channels \cite{ATLASVHCharmConstraint1,ATLASVHCharmConstraint2,CMSVHCharmConstraint1,CMSVHCharmConstraint2}. An alternative approach to these measurements consists of targeting events in which a Higgs boson is produced in association with a jet that is initiated by a charm quark, $pp\rightarrow H+c$ \cite{HiggsCharmPheno1,HiggsCharmPheno2,HiggsCharmPheno3}, henceforth referred to as the cH process. This allows for the Higgs boson to be reconstructed through a clean decay mode, such as $H\rightarrow ZZ^*\rightarrow 4\mu$, while requiring only one jet to be identified as coming from a c quark, as opposed to two in the traditional approach. The cH process is, however, a rare process in the SM, with an inclusive cross section of $\mathcal{O}(10^{-1})\ \text{pb}$. Recent measurements in the $H\rightarrow WW^*$ \cite{CMSHiggsWWCharm} and $H\rightarrow \gamma\gamma$ \cite{CMSHiggsGammaGammaCharm} decay channels have constrained the observed signal strength to $\mu_\mathrm{cH} < 257$ at 95\% CL. As a rare and relatively unconstrained process, the cH process is a good candidate for searches for physics beyond the standard model (BSM). Deviations from the SM prediction may be induced by the presence of vector-like quarks, resulting in an effective rescaling of the charm-Higgs coupling \cite{BSMYukawa1,BSMYukawa2}. Furthermore, within the framework of Two-Higgs-Doublet models for example, mixing between the light and heavy scalar fields can alter the phenomenology of the couplings of the Higgs boson to the first and second generation quarks \cite{BSMYukawaScalars1, BSMYukawaScalars2}. This can affect observables related to $H+\textrm{jet}$ processes due to insertions of additional heavy quark and scalar states at the one-loop level, in turn modifying the effective chromomagnetic dipole moments of the quarks that contribute in these processes \cite{CMDMLightQuarks1}. The sensitivity to new physics can be characterized in a model-independent way using an effective field theory (EFT) approach, where the SM is extended with operators of dimension $d > 4$. These additional terms may parametrize the effects of new particles at an energy scale $\Lambda$ that is higher than the scale that is experimentally accessible (at, for example, the LHC), while remaining agnostic about the nature of the new particles. In this work, we study how hypothetical deviations in the Higgs-charm coupling may be interpreted as arising from new physics at an energy scale $\Lambda$. Using simulated LHC proton-proton collisions in the CMS detector as an example, we demonstrate for the first time the sensitivity of the cH process to dimension-six operators.\\
\\
In Section~\ref{sec:dimsixoperators}, we give an overview of the leading-order (LO) contributions to the cH process, as well as the dimension-six EFT operators relevant to the process. In Section \ref{sec:validity}, we address the validity of the EFT interpretation. Following this, in Section \ref{sec:events}, the aspects related to the simulation, selection, and reconstruction of physics objects in the cH process in the $ZZ^*\rightarrow 4\mu$ final state are discussed in detail. With the obtained selection of cH candidate events, Section \ref{sec:observables} gives an overview of the effects of the dimension-six operators on relevant observables. In Section \ref{sec:constraints}, these observables are used in a fit resulting in expected constraints on the associated Wilson coefficients. Finally, in Section \ref{sec:summary}, we summarize the obtained results and provide an outlook to the future.

\section{Dimension-six SMEFT operators relevant to cH event production}
\label{sec:dimsixoperators}
In the standard model effective field theory (SMEFT) \cite{WarsawBasis} approach, the SM Lagrangian is extended with operators $\widehat{O}_i$ of mass dimension $d > 4$,
\begin{equation}
\label{eq:smeft}
    \mathcal{L} = \mathcal{L}_{SM} + \sum_{d>4} \sum_{i} \frac{c_i}{\Lambda^{d - 4}} \widehat{O}_i^{(d)}\,,
\end{equation}
where we have defined the Wilson coefficients $c_i$, which are the parameters of the SMEFT model, and the energy scale $\Lambda$, referred to as the energy scale of new physics. The operators are constructed out of different combinations of SM fields, resulting in terms that respect the gauge symmetries of the SM Lagrangian. In this work, we only consider operators of dimension $d = 6$ in the Warsaw basis~\cite{WarsawBasis} that affect the cH process at LO, thereby setting the Wilson coefficients of other operators to zero. Operators of odd dimension, such as the dimension $d=5$ Weinberg operator, are known to lead to baryon or lepton number violation~\cite{SMEFTOddOps}, and are therefore not considered in this study. The contribution from a single operator with Wilson coefficient $c$ to a cross section is
\begin{equation}
\label{eq:xsec}
    \sigma = \sigma_{\text{SM}} + \frac{c}{\Lambda^2} \delta_{\text{int}} + \frac{c^2}{\Lambda^4} \sigma_{\text{EFT}}\,,
\end{equation}
where $\sigma_{\text{SM}}$ is the SM cross section computed from matrix-element amplitudes featuring only SM interaction vertices, $\sigma_{\text{EFT}}$ is the cross section resulting from matrix elements featuring exactly one EFT vertex insertion, and $\delta_{\text{int}}$ contains the interference between the purely SM and EFT amplitudes. The term proportional to $c/\Lambda^2$ is referred to as the interference term, while the term proportional to $c^2/\Lambda^4$ is known as the quadratic term. While operators of higher dimension $d \geq 8$ could contribute to the cH process, Eq. \eqref{eq:smeft} implies that their effect on the cross section is suppressed by at least a factor $\Lambda^{-4}$ in the interference term and $\Lambda^{-8}$ in the quadratic term. Therefore, the effect of these operators are assumed to not impact this study. In the following sections, we briefly introduce the SMEFT operators relevant to the cH process. For each of the operators discussed, we only consider the CP-conserving variants as the targeted observables are not sensitive to CP-violating effects.
\begin{figure}
    \centering
    \includegraphics[width=\linewidth]{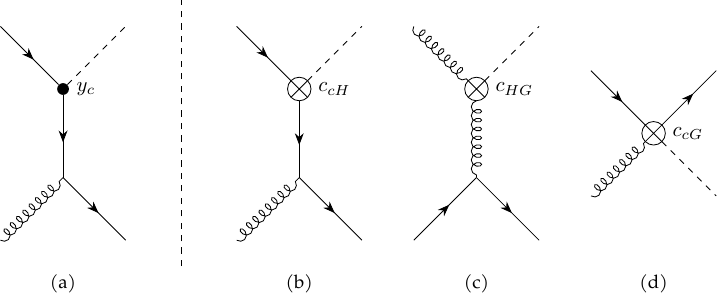}
    \caption{Example contributions to cH production at LO, mediated by the SM Yukawa vertex (a) and by vertices associated with dimension-six operators in the SMEFT (b,c,d).}
    \label{fig:diagrams}
\end{figure}
In figure \ref{fig:diagrams}, example Feynman diagrams are displayed illustrating the SM and SMEFT contributions to cH production considered in this study.
\subsection{The chromomagnetic dipole operator}
The first operator to be considered is $\hat{O}_{cG}$, defined as
\begin{equation}
\label{eq:cmdoperator}
    \hat{O}_{cG} = \left(\bar{q}_2 \sigma^{\mu \nu} T^a \textrm{c}\right) \tilde{H} G_{\mu \nu}^a\,,
\end{equation}
with the left-handed second-generation quark doublet $q_2$, the right-handed charm field $\textrm{c}$, the conjugate scalar doublet $\widetilde{H} = i\sigma_2 H^*$ with $\sigma_2$ the second Pauli matrix, the QCD field strength $G^a_{\mu\nu}$ with generators $T^a$, and finally $\sigma^{\mu\nu} = \frac{i}{2} [\gamma^\mu, \gamma^\nu]$ with $\gamma^\mu$ the Dirac matrices. This operator is also known as the chromomagnetic dipole (CMD) operator for the charm quark, as it generates tree level contributions to the charm anomalous chromomagnetic dipole moment $\hat{\mu}_c$ of the form $\left(\bar{\textrm{c}} \sigma^{\mu\nu} \hat{\mu}_\textrm{c} \textrm{c} \right) T^a G^{a}_{\mu\nu}$. The charm CMD operator introduces a four-point interaction of the type $\textrm{c}_L\textrm{c}_RHG$ resulting in diagram (d) in Fig.~\ref{fig:diagrams}. Note that the CMD operator mixes left-handed ($L$) and right-handed ($R$) chiral fields. In the SM, this type of tree level mixing is unique to the Yukawa interactions $\propto \bar{f_L} H f_R$ with the fermion fields $f$. An important consequence of this chiral structure is that interference with the SM is generally suppressed by the charm mass $m_\textrm{c}$, which determines the magnitude of mixing of left and right chiral charm states, and thus severely limits the sensitivity of many SM processes to $c_{cG}$. In fact, in Refs. \cite{BottomCMDStudy} and \cite{GeneralCMDStudy} it is argued that the CMD operators for the $\textrm{b}$, $\textrm{c}$, and light quarks cannot be constrained by dijet and flavor-blind Higgs boson production analyses as a result of the chirality-suppressed interference terms. Therefore, cH production is an ideal (and perhaps the only) candidate to constrain $c_{cG}$. The reason for this is that the interference term between diagram (a) and diagram (d) in Fig.~\ref{fig:diagrams} is not suppressed by an additional factor of $m_c$ and the terms quadratic in $c_{cG}$ are competing with a much smaller $\sigma_\textrm{SM}$.

\subsection{The Yukawa operator}
Another dimension-six operator affecting cH production is given by
\begin{equation}
\label{eq:yukawa}
\hat{O}_{cH} = \left( H^\dagger H \right) \left( \bar{q}_2 \tilde{H} c \right)\,.
\end{equation}
Its structure is analogous to the SM Yukawa operator with two additional instances of the SM Higgs doublets. For this reason, we will refer to $\hat{O}_{cH}$ as the dimension-six charm Yukawa operator. When the two Higgs fields in Eq.~\eqref{eq:yukawa} acquire their vacuum expectation values (VEVs), the resulting term effectively rescales the SM charm Yukawa coupling $y_c$. In Fig. \ref{fig:diagrams}, diagram (b) features an insertion of a $ccH$ vertex generated by the $\hat{O}_{cH}$ operator. We can formalize this connection between the dimension-six Yukawa operator and the SM operator by defining the coupling modifier $\kappa_c = y_c / y_c^{SM}$, such that
\begin{equation}
\label{eq:kappac}
\kappa_c = 1 - \frac{v^3}{m_c} \frac{c_{cH}}{\Lambda^2}\,,
\end{equation}
where $v$ is the VEV and $c_{cH}$ is the Wilson coefficient of the $d=6$ charm Yukawa operator. Any result quoting limits on $\kappa_c$ can therefore be translated into individual limits on $c_{cH}$. The strongest constraint on $\kappa_c$ currently comes from the combination of the $ttH(c\bar{c})$ and $VH(c\bar{c})$ analyses with the CMS experiment \cite{CMSVHCharmConstraint2} using the full LHC Run 2 dataset, resulting in an observed limit of $|\kappa_c| < 3.5$. Using Eq.~\eqref{eq:kappac}, this results in an individual limit on the dimension-six charm Yukawa coefficient of $|c_{cH}/\Lambda^2| < 2.1\times 10^{-2}$ TeV$^{-2}$. Upper limits on $\kappa_c$ from existing analyses of cH production are an order of magnitude larger \cite{CMSHiggsWWCharm,CMSHiggsGammaGammaCharm}, therefore we do not expect to obtain constraints on $\tilde{c}_{cH}$ that are competitive with this result.

\subsection{The Higgs-gluon operator}
The Higgs-gluon operator is given by
\begin{equation}
\label{eq:higgsgluon}
    \hat{O}_{HG} = \left( H^\dagger H \right) G^{a\mu\nu}G_{\mu\nu}^a\,.
\end{equation}
The operator generates contact interactions between the Higgs field and the gluon fields, as demonstrated in diagram (c) in Fig.~\ref{fig:diagrams}. It has traditionally been studied in the context of Higgs boson production rates, as it generates a tree-level contribution to the gluon fusion Higgs boson production mode. As a result, individual limits on the Higgs-gluon operator $\hat{O}_{HG}$ already tightly constrain the Wilson coefficient $c_{HG}$. For example, in the combined measurements of Higgs boson and decays with the CMS experiment \cite{CMSHiggsCombinedMeasurements}, $c_{HG}$ has been bounded to the interval $[-0.0039, 0.0042]$ at 95\% CL.

\subsection{Contributions to the cH cross section}
Here, we estimate the relative sizes of the terms in 
Eq.~\eqref{eq:xsec} for the operators considered in this study. To this end, we use \texttt{MadGraph\_aMC@NLO} \cite{MadGraphMain} to generate the matrix elements for the cH process using the Feynman rules provided by the \texttt{SMEFTsim} model \cite{SMEFTsim1,SMEFTsim2}, which features a complete implementation of the dimension-six operators in the Warsaw basis \cite{WarsawBasis}. We impose a generator-level requirement on the transverse momentum of the produced charm quark of $p_{T} > 20$ GeV to replicate an analysis-like selection. The cross section may be normalized to the SM cross section, to define the signal strength $\mu_{cH}$ of the cH process:
\begin{equation}
\label{eq:sigstrength}
    \mu_{cH} = \frac{\sigma}{\sigma_{\text{SM}}} = \left( 1 + \frac{c}{\Lambda^2}\mu_{\text{int}} + \frac{c^2}{\Lambda^4} \mu_{\text{quad}}\right),
\end{equation}
where $\sigma_{\text{SM}} = 0.056$ pb at LO. In the above expression, $\mu_{\text{int}}$ and $\mu_{\text{quad}}$ represent the contributions from the linear and quadratic terms to the signal strength $\mu_{cH}$. In these calculations, we set the scale to $\Lambda = 1$ TeV and estimate $\mu_\textrm{int}$ and $\mu_\textrm{quad}$ by fitting Eq. \eqref{eq:sigstrength} to the $(c, \mu_{cH})$ pairs resulting from the calculation of the matrix element generator. The results are shown in Table~\ref{table:xsec}. For reference, we use $\mu_{cH} < 257$ at 95\% CL as the best limit \cite{CMSHiggsWWCharm} on the cH signal strength to illustrate the bounds of the Wilson coefficients that these imply. 
\begin{table}[h!]
 \centering
  \caption{The relative sizes of the interference and quadratic contributions to the Hc cross section for the operators considered in this study.}
  \label{table:xsec}
  \begin{tabular}{llccc}
    \toprule
     Operator & Wilson coefficient & $\mu_{\text{int}}$ & $\mu_{\text{quad}}$ & Region where $\mu_{cH} < 257$\\
    \midrule
    $\widehat{O}_{cG}$ & $c_{cG}$ & 2.2 & 55.8 & $c_{cG} \in [-2.16,2.12]$\\
    $\widehat{O}_{cH}$ & $c_{cH}$ & -16.6 & 69.0 & $c_{cH} \in [-1.81,2.05]$\\
    $\widehat{O}_{HG}$ & $c_{HG}$ & 0.5 & 1608.6 & $c_{HG} \in [-0.40, 0.40]$\\
    \bottomrule
  \end{tabular}
\end{table}
For values of $|c|>1$, we can see that the interference terms will generally be much smaller compared to the quadratic terms, especially for $c_{cG}$ and $c_{HG}$. For the $\hat{O}_{cH}$ operator, we observe that $\mu_\textrm{int}$ is larger in magnitude with respect to the other operators, likely owing to the similarity with the SM Higgs-charm Yukawa operator. As expected, for values of $|c|>1$, the quadratic terms are dominant for all three operators, as these terms are not suppressed by the small SM charm Yukawa coupling $y_c$.

\section{Validity of SMEFT interpretation}
\label{sec:validity}
In practice, the SMEFT Lagrangian in Eq.~\eqref{eq:smeft} is often truncated at some dimension $d$. The resulting model is known not to be well-behaved in the UV regime due to not being renormalizable. The $\Lambda$ parameter thus represents the scale at which the predictions from the SMEFT model are no longer reliable. Beyond this scale, resonances associated with new particles are assumed to appear. The value of $\Lambda$ is arbitrary, but nonetheless critical to the predictions of the SMEFT. As observables such as the cross section in Eq.~\eqref{eq:xsec} are degenerate in the values of $\Lambda$ and the Wilson coefficients $c_i$, choosing a value of $\Lambda$ which is too small may result in unrealistic limits on the $c_i$, as the process under study may be characterized by energy scales which exceed $\Lambda$. Instead, we can absorb $\Lambda$ into the definition of the Wilson coefficients, $\tilde{c}_i = c_i / \Lambda^2$ and parameterize our limits on $\tilde{c}_i$ as a function of a cutoff value $M_\mathrm{cut}$, which represents the energy scale that characterizes the process under study. \\
\\
As discussed in Ref.~\cite{EFTValidity}, another requirement for a valid SMEFT interpretation is that the obtained limits on $\tilde{c}_i(M_\mathrm{cut})$ are sensible as an interpretation of the low-energy limit of a UV-complete theory. Naturally, verifying this statement requires us to make assumptions on the structure of the UV models, thereby losing the generalized treatment that is associated with the SMEFT approach. To quantify the validity of the SMEFT interpretation, we will instead resort to a broad and simplified scenario, where we assume the BSM model only introduces one additional new-physics coupling $g_*$ at some scale $\Lambda$. We can use dimensional power-counting rules to characterize the dependence of the Wilson coefficients on the new-physics coupling $g_*$. In this work, we use the counting scheme of naive dimensional analysis (NDA) \cite{NaieveDimAnalysis},
\begin{equation}
    \label{eq:eftpowercounting}
\mathcal{L}_{\mathrm{eff}}=\frac{\Lambda^4}{g_*^2} \mathcal{L}\left(\frac{D_\mu}{\Lambda}, \frac{g_* H}{\Lambda}, \frac{g_{*} f_{L, R}}{\Lambda^{3 / 2}}, \frac{g F_{\mu \nu}}{\Lambda^2}\right)\,,
\end{equation}
where $D$, $H$, $f$ and $F$ represent instances of the covariant derivative, scalar doublet, fermion field and SM gauge field, respectively. The NDA scheme describes with which powers of $g_*$ and $\Lambda$ the SM fields appear in the EFT operators under study. Note that the SM gauge fields $F_{\mu\nu}$ appear only with their respective couplings $g$, as the corresponding gauge symmetries are assumed to hold at energies above $\Lambda$. Applying these power-counting rules to the CMD operator in Eq.~\eqref{eq:cmdoperator} yields the following expression:
\begin{align*}
    \left(\bar{q} \sigma T \textrm{c}\right) \tilde{H} G \longrightarrow&
    \,\frac{\Lambda^4}{g^2_\star} 
    \cdot 
    \left[
        \left(\frac{g_\star f_{L,R}}{\Lambda^{3/2}}\right) 
        \cdot \left(\frac{g_\star f_{L,R}}{\Lambda^{3/2}}\right) 
        \cdot \left(\frac{g_\star H}{\Lambda}\right) 
        \cdot \left(\frac{g_s G}{\Lambda^2}\right)
    \right]\\
    &= \frac{g_\star g_s}{\Lambda^2} \left( f_{L,R}f_{L,R}HG \right)\,.
\end{align*}
Therefore, in the low-energy limit of the simple BSM theory, the coupling of the CMD operator can be interpreted as $g_\star g_s / \Lambda^2$. In SMEFT, this coupling is equivalent to $\tilde{c}_{cG}$, such that
\begin{equation*}
  \left| \tilde{c}_{cG} \right| = \frac{g_\star g_s}{\Lambda^2} < \frac{g_\star g_s}{M_{\mathrm{cut}}^2}
  \Rightarrow \left| \tilde{c}_{cG} \right| M_{\mathrm{cut}}^2 < g_\star g_s\,.
\end{equation*}
Thus, given some value of $M_\mathrm{cut}$, it follows that the BSM coupling $g_*$ cannot be arbitrarily large as this would spoil the perturbativity of the model. This is typically expressed through the requirement that $g_*$ is roughly $\leq 4\pi$. As a result, our perturbativity requirement for the CMD operator now becomes
\begin{equation}
  \label{eq:perturbativityCMD}
  \frac{\left| \tilde{c}_{cG} \right| M_{\mathrm{cut}}^2}{4\pi g_s} \lesssim \epsilon\,,
\end{equation}
where choosing $\epsilon = 1$ corresponds to loosely satisfying the perturbativity requirement, and smaller values correspond to more strict interpretations. 
\begin{figure}[t]
\centering
\begin{subfigure}{0.48\textwidth}
\centering
\includegraphics[width=\linewidth]{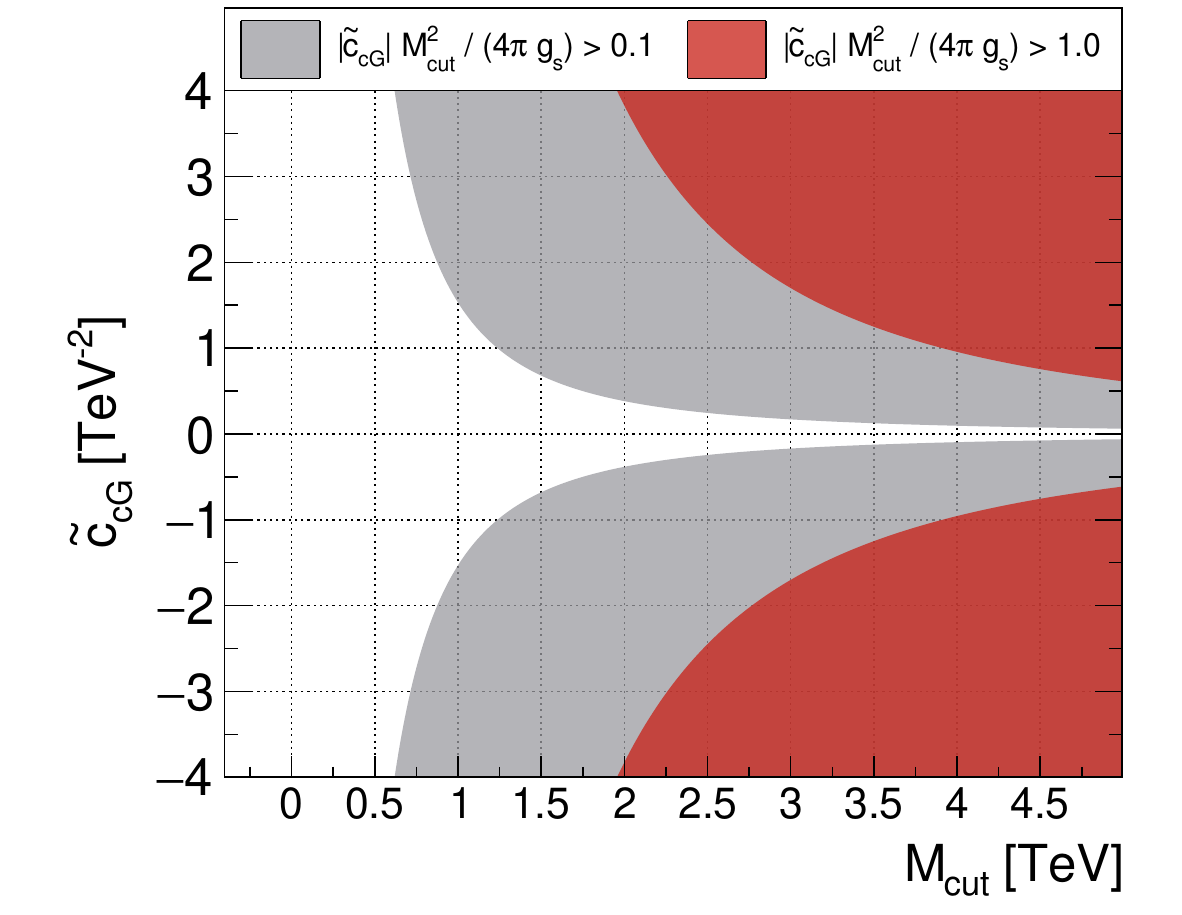}
\caption{}
\label{fig:validityCMD}
\end{subfigure}
\hfill
\begin{subfigure}{0.48\textwidth}
\centering
\includegraphics[width=\linewidth]{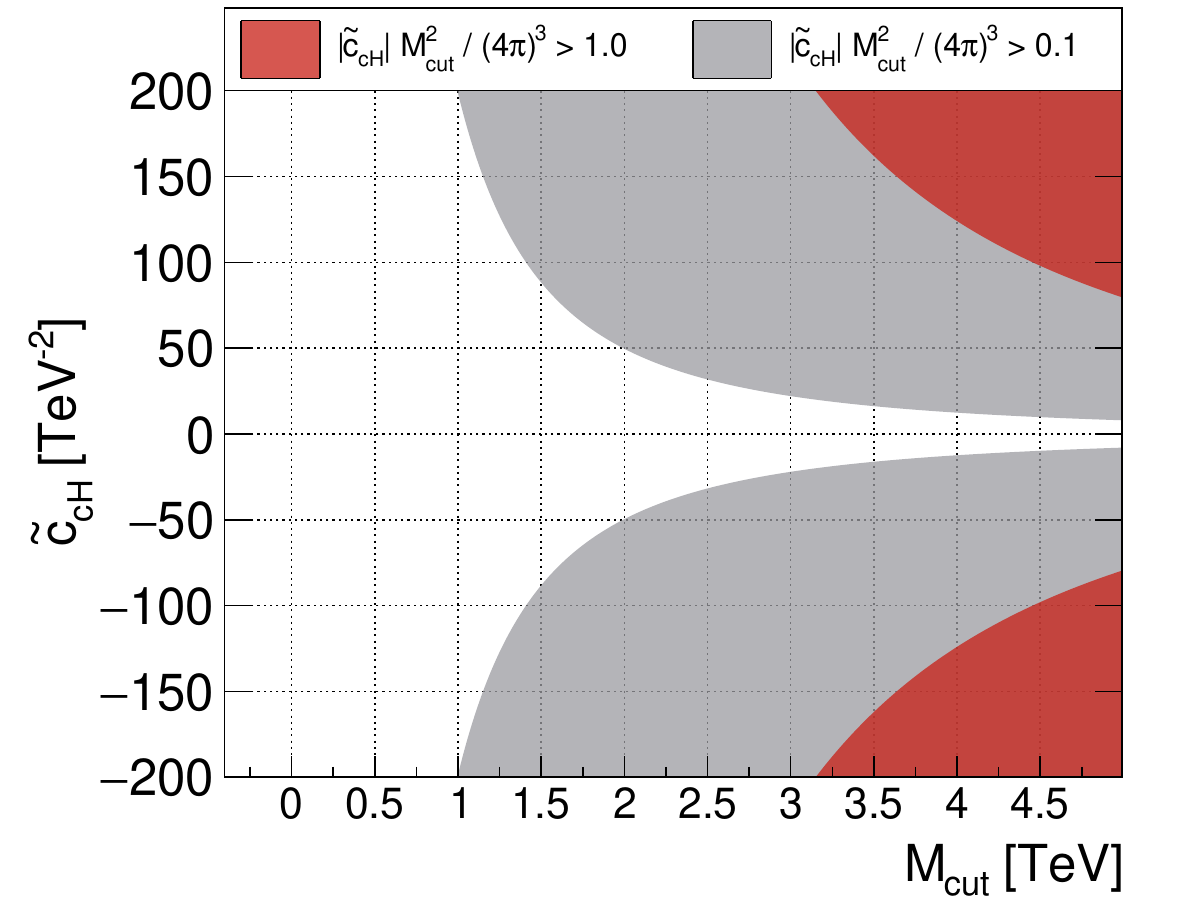}
\caption{}
\label{fig:validityYukawa}
\end{subfigure}
\caption{Perturbativity bounds for $\tilde{c}_{cG}$ (a) and $\tilde{c}_{cH}$ (b) as a function of $M_{\textrm{cut}}$ in TeV.}
\label{fig:validityboth}
\end{figure}
These bounds are shown in Fig.~\ref{fig:validityCMD} for $\epsilon = 1$ and $\epsilon = 0.1$. We remark that, already for $M_\mathrm{cut} = 1.0$ TeV, the coefficient of the CMD operator needs to be limited to the interval $[-1.5, 1.5]$ TeV$^{-2}$ to avoid spoiling the perturbativity of the simple BSM theory. For the  Yukawa operator given by Eq.~\eqref{eq:yukawa}, a similar procedure yields
\begin{equation}
\label{eq:valYuk}
  \frac{|\tilde{c}_{cH}| M^2_{\textrm{cut}}}{\left( 4 \pi \right)^3} \lesssim \epsilon\,.
\end{equation}
As shown in Fig.~\ref{fig:validityYukawa}, the additional factors of $4\pi$ in the denominator of \eqref{eq:valYuk} result in the validity constraints on $\tilde{c}_{cH}$ being generally much more relaxed compared to those on $\tilde{c}_{cG}$. For example, using $M_\mathrm{cut} = 1.0$ TeV, we now have that $\tilde{c}_{cH} \in [-200, 200]$ TeV$^{-2}$. In this work, the invariant mass of the H+jet system $m_{\textrm{H+jet}}$ will represent the energy scale characteristic to the cH process. This variable will therefore be employed as the $M_\mathrm{cut}$ parameter to interpret the validity of the obtained limits on the operators considered in this study under the assumption of a UV-completion featuring a single new-physics coupling.

\section{Event simulation, selection, and reconstruction}
\label{sec:events}
In the following sections, the sensitivity of the cH process to the CMD and Yukawa operators is studied by targeting cH($ZZ^*\rightarrow 4\mu$) final states. To this end, we have produced a sample of simulated cH($ZZ^*\rightarrow 4\mu$) events. Specifically, proton-proton collisions at a center-of-mass energy of 13 TeV at the LHC are simulated using the Run-2 configuration of the CMS detector.

\subsection{Event simulation}
\label{sec:simulation}
The matrix elements and parton-level events for the cH process are generated at LO with \texttt{MadGraph\_aMC\@NLO} \cite{MadGraphMain}. The Feynman rules associated with the operators considered in this work are implemented by the \texttt{SMEFTsim} \cite{SMEFTsim1, SMEFTsim2} UFO model, which facilitates the calculation of matrix elements featuring insertions of SMEFT interaction vertices. The parameters of the model are restricted by setting the values of all Wilson coefficients to zero, apart from the CMD and Yukawa operators which are set to non-zero values. Differential distributions associated with various SMEFT scenarios are obtained using MadGraph's built-in reweighting technique \cite{MadGraphReweight}. The parton-level events are showered using \texttt{Pythia8} \cite{Pythia8}. The detector response and object reconstruction are simulated with \texttt{Delphes} \cite{Delphes} using a modified version of the default CMS detector card to include a parameterization of the DeepJet c-tagging  algorithm \cite{DeepJet} and the simulation of misidentified muons. For the background processes relevant to the $4\mu+1$ jet signal topology, we make use of several simulated datasets from the CMS OpenData platform \cite{OpenDataSample0,OpenDataSample1,OpenDataSample2,OpenDataSample3,OpenDataSample4,OpenDataSample5,OpenDataSample6,OpenDataSample7,OpenDataSample8,CMSOpenDataDY}. The generator-level objects of these datasets are used as input to Delphes to obtain a set of background samples, the detector response of which is simulated with the same setup as the signal sample. The irreducible background is composed of the processes with four prompt muons in the final state. They include the dominant Higgs boson production modes, including gluon fusion ($ggH$), vector boson fusion (VBF), vector boson associated production ($VH$), production in association with top quarks ($ttH$ and $tqH$), Higgs boson production in association with a $b$ jet ($bH$) mediated by Yukawa interactions, and $ZZ$ production via a quark-antiquark initial state and via gluon fusion. 
The reducible background consists of the Drell-Yan process ($Z/\gamma^*+\text{jets}$) with $Z/\gamma^*\rightarrow \mu^+\mu^-$, and $t\bar{t}+\text{jets}$ production, as these processes could involve two muon candidates originating from non-prompt or misreconstructed muons. All processes are generated at next-to-LO (NLO) in perturbative QCD, with the exception of $ZZ$ production induced by pairs of gluons, which is generated at leading order (LO) in perturbative QCD. 
To account for significant differences in cross section of $ZZ$ production between the used simulation samples and known higher-order QCD corrections, next-to-NLO/NLO $K$ factors are applied for the $gg\rightarrow ZZ$ and $qq\rightarrow ZZ $ simulations \cite{qqZZkFactor}\cite{ggZZkFactor}.

\subsection{Event selection and reconstruction}
\label{sec:selectionreconstruction}
\begin{figure}[t]
\centering
\includegraphics[width=0.5\linewidth]{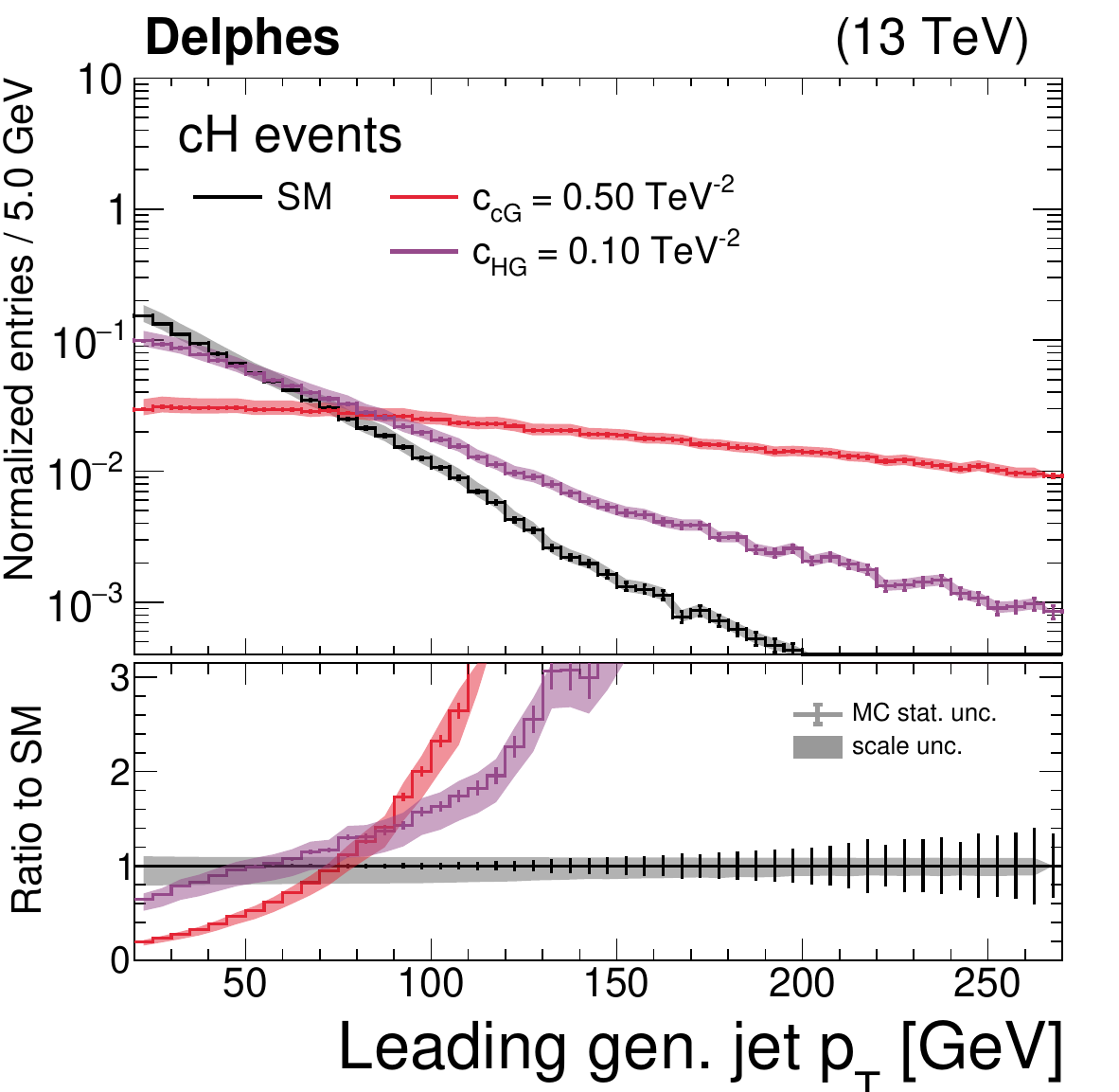}
\caption{Transverse momentum of the leading jet for SM cH($ZZ^*\rightarrow 4\mu$) events as well as for scenarios corresponding to non-zero values for the $c_{cG}$ and $c_{HG}$ Wilson coefficients. The histograms are normalized to unity to allow for a direct comparison of the shapes. Both operators result in a significant enhancement of the tail of the leading jet $p_{T}$ distribution in simulated events.}
\label{fig:genPtCMD}
\end{figure}

This section details the reconstruction algorithm used to reconstruct cH($ZZ^*\rightarrow 4\mu$) event candidates. First, an initial object selection is performed, which consists of selecting a set of muon and jet candidates per event. The criteria applied in this selection is shown in Table~\ref{table:preSelection}. The effect of the CMS muon identification criteria at the tight working point \cite{CMSMuonID} is mimicked by sampling from an efficiency distribution that reflects the efficiency of said criteria. This distribution is dependent on the muon transverse momentum and pseudorapidity and is derived using the simulated Drell-Yan samples \cite{CMSOpenDataDY} by determining the efficiency with which muons produced from Z boson decays are identified with this set of criteria, in the CMS detector. The muon isolation criteria that are applied correspond to the same ones used by the CMS collaboration \cite{CMSMuonID}. For the identification of jets originating from pileup, a cut-based pileup identification technique is applied using the \texttt{Delphes} built-in \texttt{PileUpJetID} module based on the discriminating variables defined in Ref. \cite{PileupCutBased}. \\
\\
\begin{table}[htbp]
\centering
\begin{tabular}{l l}
\hline
\textbf{Object} & \textbf{Selection criteria}\\
\hline
Muons & $p_T > 5$ GeV \\
& $\vert \eta \vert < 2.4$ \\
& Muon identification criteria\\ 
& Muon isolation criteria \\
\hline
Jets & $p_T > 20$ GeV \\
& $\vert \eta \vert < 2.5$ \\
& Jet pileup identification criteria \\
\hline
\end{tabular}
\caption{Object selection criteria for muons and jets, used to reconstruct cH($ZZ^*\rightarrow 4\mu$) event candidates.}
\label{table:preSelection}
\end{table}
\\
Following the initial object selection, four-muon events are selected for the reconstruction of a Higgs boson candidate. The strategy applied here is inspired by the one presented in Ref.~\cite{HIG19-001}. To reconstruct a Higgs boson candidate, events with exactly four muons and a net charge of zero are selected. These are all required to be well-separated by applying a $\Delta R > 0.02$ criteria between any two muons to avoid erroneous, duplicate reconstructions of muons. The muon (sub)leading in transverse momentum is required to satisfy $p_{T} > 20$ (10) GeV. From these four muons, Z boson candidates are reconstructed by pairing opposite-sign muons. The invariant dimuon mass of all of these pairings is required to satisfy $m_{\mu\mu} > 4$ GeV, to ensure contributions from low-mass resonances such as $J/\psi$ are rejected. Subsequently, the Z boson candidate with a minimal difference of the dimuon mass to the world-average value of the Z boson mass $m_Z = 91.19$ GeV \cite{PDG} is taken as the on-shell $Z_1$ boson candidate. This candidate is required to satisfy $40$ GeV$ < m_{Z_{1}} < 120$ GeV.  The remaining candidate is taken as the $Z_2$ boson candidate, which may be considerably off-shell and thus a relaxed requirement of  $12$ GeV$ < m_{Z_{2}} < 120$ GeV is applied. Finally, the $Z_1$ and $Z_2$ boson candidates are used to construct a Higgs boson candidate. It is required that the four-muon mass of the Higgs boson candidate satisfies $m_{4\mu} > 70$ GeV, since contributions from Higgs boson production below this value become negligible.  \\
\\
After the Higgs boson reconstruction is performed, the jet leading in transverse momentum is picked to complete the cH($ZZ^*\rightarrow 4\mu$) event candidate reconstruction. This choice is particularly suitable when considering the CMD operator, due to the high transverse momentum contribution to the charm quark that the CMD operator introduces for the cH process, as can be seen in Fig.~\ref{fig:genPtCMD}.

\section{Effect of SMEFT operators on observables}
\label{sec:observables}
\begin{figure}[t]
\centering
\begin{subfigure}{0.49\textwidth}
\centering
\includegraphics[width=\linewidth]{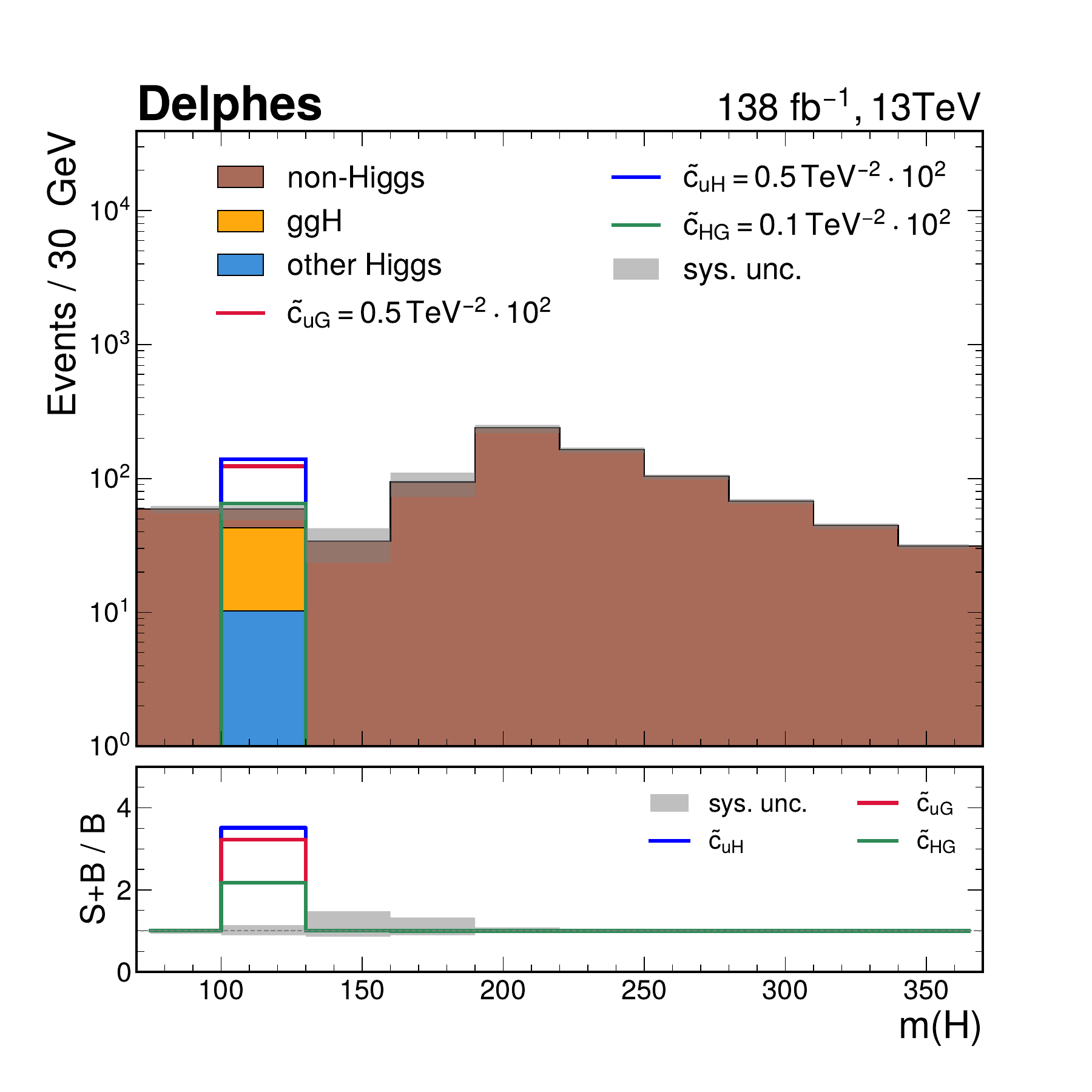}
\caption{Mass of the reconstructed Higgs boson candidate.}
\end{subfigure}
\hfill
\begin{subfigure}{0.49\textwidth}
\centering
\includegraphics[width=\linewidth]{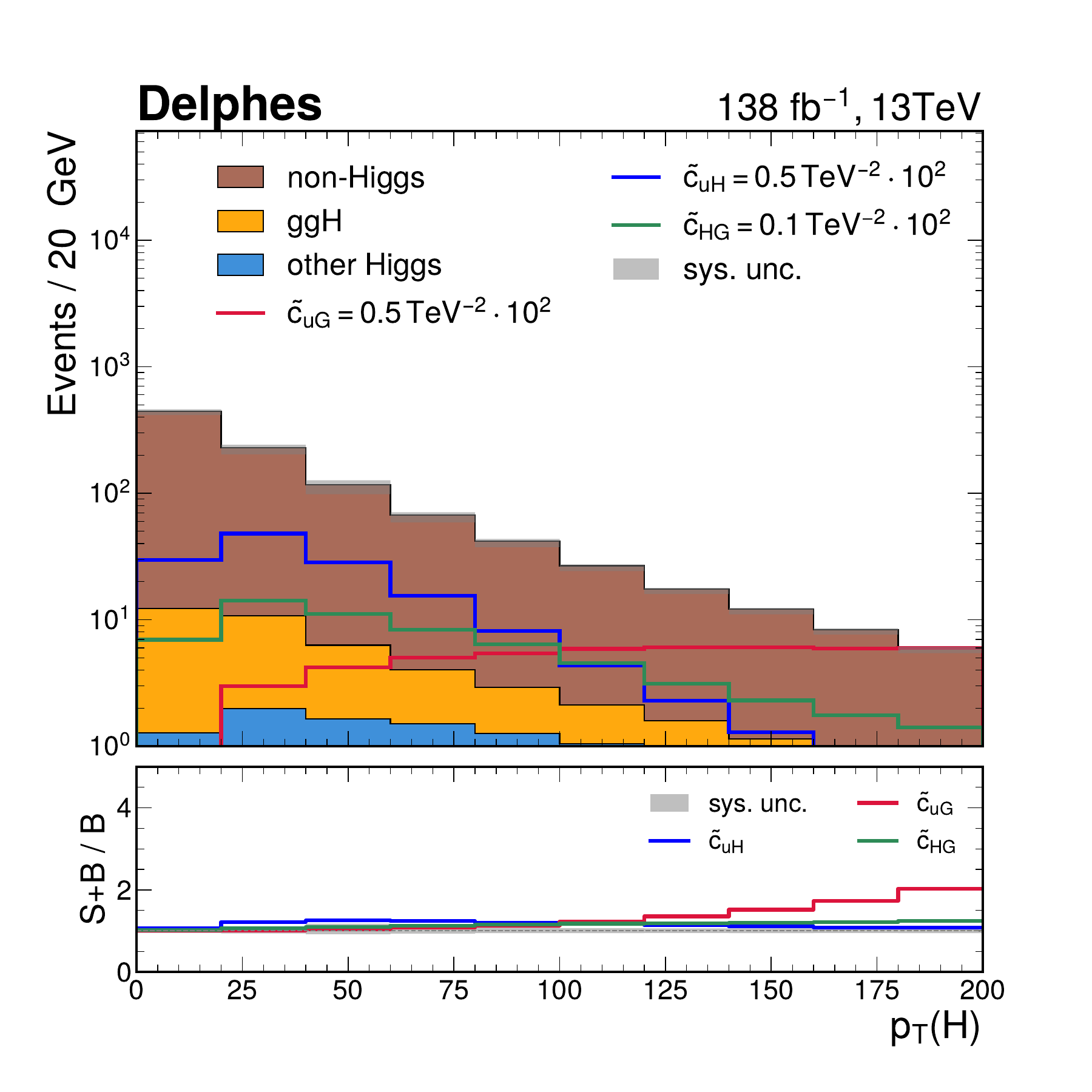}
\caption{Transverse momentum of the Higgs boson candidate.}
\end{subfigure}
\begin{subfigure}{0.49\textwidth}
\centering
\includegraphics[width=\linewidth]{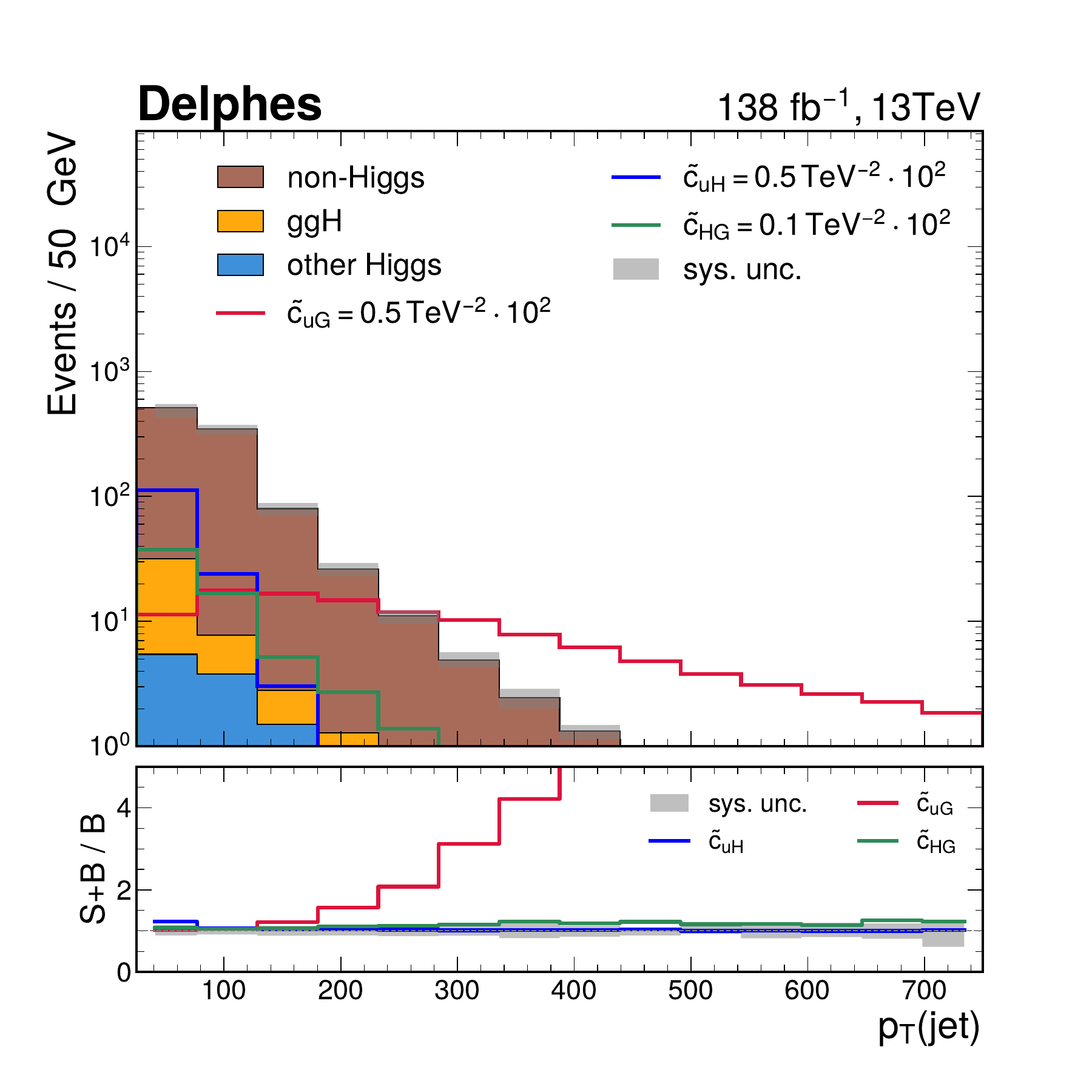}
\caption{Transverse momentum of the selected jet candidate.}
\label{fig:LeadingJetPt}
\end{subfigure}
\hfill
\begin{subfigure}{0.49\textwidth}
\centering
\includegraphics[width=\linewidth]{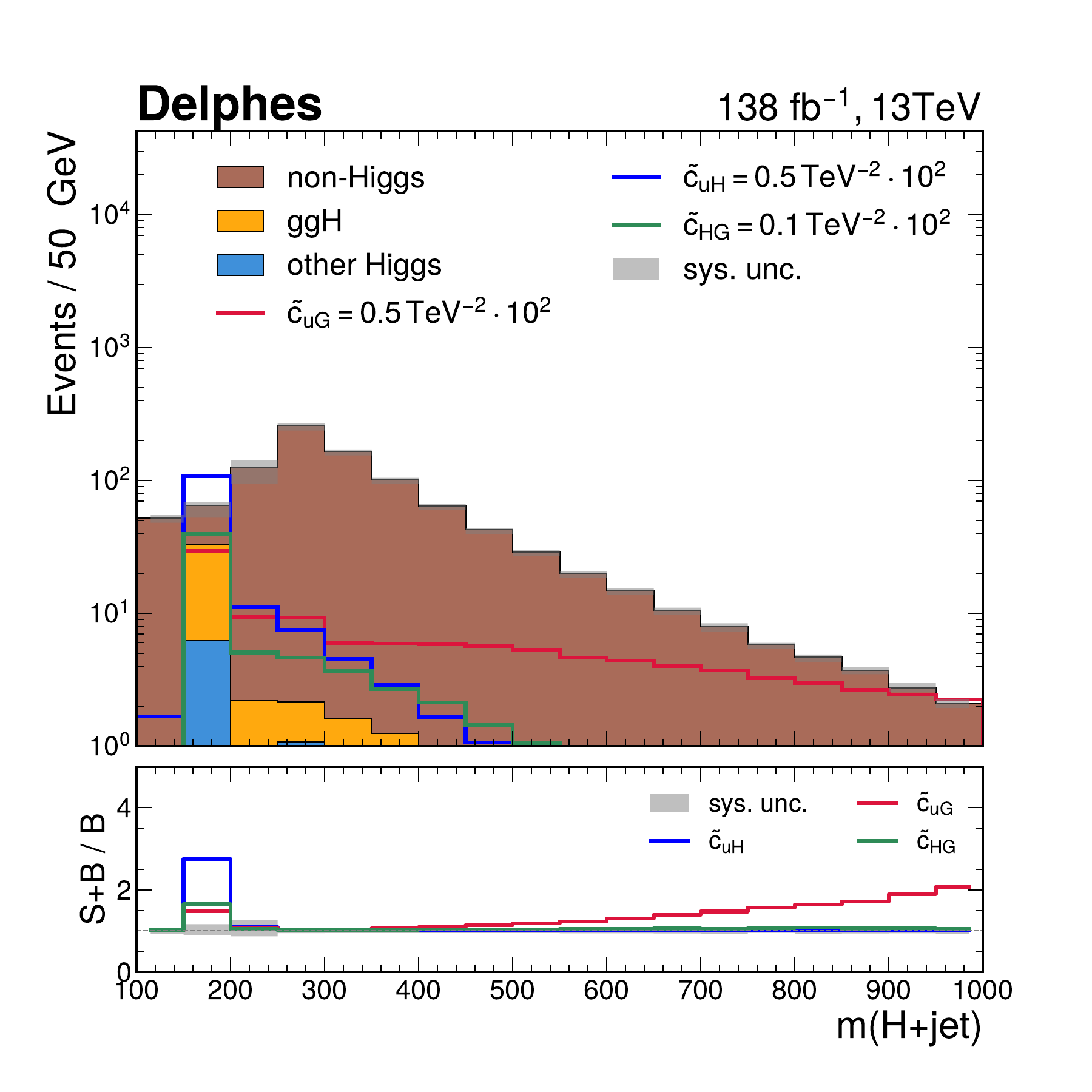}
\caption{Mass of the combined Higgs boson candidate and jet candidate system.}
\label{fig:HJetMass}
\end{subfigure}
\caption{Higgs boson and jet candidate observable distributions, showing both background processes and the signal stemming from the $\hat{O}_{cG}$, $\hat{O}_{cH}$ operators and $\hat{O}_{HG}$ operators. Here, these are simulated with nominal coefficients $\tilde{c}_{cG} = \tilde{c}_{cH} = 0.5$~TeV$^{-2}$ and $\tilde{c}_{HG} = 0.1$~TeV$^{-2}$. The simulation is shown for an integrated luminosity of 138 fb$^{-1}$ at a center-of-mass energy of 13 TeV. The signal contributions are scaled by a factor of of
hundred to improve their visibility.}
\label{fig:Observables}
\end{figure}

 Distributions of observables derived from applying the reconstruction detailed in Section~\ref{sec:selectionreconstruction} on the discussed simulated SMEFT and background process datasets are shown in Fig.~\ref{fig:Observables}. Contributions from the $\hat{O}_{cG}$ and $\hat{O}_{cH}$ operators, the effects of which are shown for example choices of $\tilde{c}_{cG} = 0.5$~$\mathrm{TeV}^{-2}$ and $\tilde{c}_{HG} = 0.1$~$\mathrm{TeV}^{-2}$, clearly produce an enhancement of the Higgs boson resonance at $m(H)$ = 125 GeV. The signal arising from the $\hat{O}_{cG}$ operator on the other hand, produces a significant enhancement in the high transverse momentum region of both the jet and Higgs boson candidates, with the effect being more pronounced for the former. This behavior, which originates from the introduction of the four-point interaction associated with $\hat{O}_{cG}$, is also reflected in the mass of the combined Higgs boson and jet candidate system, $m_\mathrm{H+\mathrm{jet}}$. Similarly, some enhancement is observed in the high transverse momentum region of the jet candidate for the signal associated with $\hat{O}_{HG}$, though this effect is comparatively small with respect to the $\hat{O}_{cG}$ signal. 

\section{Constraining Wilson coefficients in cH events}
\label{sec:constraints}
This section details the statistical evaluation that is applied to the selected cH candidate events to derive expected 95\% CL upper limits (henceforth referred to as \textit{upper limits}) on the Wilson coefficients $c_{cG}$ and $c_{cH}$ via the asymptotic CL$_{\mathrm{s}}$ method~\cite{CLs}. These upper limits are derived for an integrated luminosity of 138 fb$^{-1}$, as also in Refs. \cite{CMSHiggsWWCharm,CMSHiggsGammaGammaCharm}, which corresponds to the dataset recorded by the CMS detector during Run-2 of the LHC. This result is also extrapolated to an integrated luminosity of 3000 fb$^{-1}$, which is anticipated at the High Luminosity-LHC (HL-LHC) \cite{HL-LHC}. 

\subsection{Statistical evaluation strategy}
\label{sec:fit}
To derive expected upper limits on $\tilde{c}_{cG}$ and $\tilde{c}_{cH}$, the \textsf{Combine} tool \cite{Combine} is used. Specifically, the profile likelihood ratio test statistic \cite{Cowan} is employed in a binned likelihood fit, using a discriminator derived from the event selection described in Section~\ref{sec:selectionreconstruction}. Given the significant enhancement of the jet transverse momentum associated with the $\hat{O}_{cG}$ operator, the transverse momentum of the jet candidate is chosen as the discriminator for this operator. To reduce the effect of background processes, particularly ZZ production, an additional Higgs boson candidate mass criterion of $m(H) \in [100, 130]$ GeV is applied. For the $\hat{O}_{cH}$ operator on the other hand, the Higgs boson candidate mass itself is chosen as the discriminator, given it is primarily associated with an enhancement in the jet-associated Higgs boson production rate. \\
\\
In a fit of the Wilson coefficient $c_i$, both the interference term linear in $c_i$, and the purely EFT term, quadratic in $c_i$, must be accounted for to produce a correct scaling of $c_i$ during the fitting process. This is as each contribution has a unique discriminator distribution shape, thus resulting in a non-trivial scaling. This is achieved by including separate simulations for the linear and quadratic terms for each operator $\hat{O}_{i}$, thus ensuring the correct scaling.

\subsection{Systematic uncertainties}
\label{sec:uncertainties}
A number of systematic uncertainty sources expected to have an impact in the measurement are accounted for during the fitting procedure by parametrizing them as nuisance parameters. These sources fall into two categories: experimental uncertainties and theoretical uncertainty sources, which are discussed below.

\subsubsection{Experimental uncertainty sources}
A number of experimental uncertainty sources are considered. For example, uncertainties associated with the calibration of the jet energy scale and jet energy resolution are estimated using a parametrization of the uncertainties reported in Ref.~\cite{JetPeformanceRun2}. While the yield-changing effect for these uncertainties are negligible, they introduce changes to the shape of the discriminator. Uncertainties associated with the calibration of the muon reconstruction efficiency are estimated using a parametrization of the uncertainties reported in Ref.~\cite{MuonEfficiencies}. This uncertainty is associated with a yield-changing effect of approximately 1--2\%. The uncertainty stemming from the limited precision with which the luminosity at the LHC is known is also considered, based on the uncertainty reported in Ref.~\cite{LumiUnc}. 
\subsubsection{Theoretical uncertainty sources}
A number of theoretical uncertainties that affect the simulation of physics processes are included. This includes an uncertainty associated with the \textsf{NNPDF3.1} PDF set \cite{NNPDF3.1} that is used in the simulation. For this, an uncertainty envelope is constructed using 100 variations of the PDF. This uncertainty introduces a yield-changing effect of approximately 6--7\% for the background processes and up to 20\% for the signal processes. The uncertainty arising from the choice of the strong coupling constant $\alpha_s$ in the PDF is additionally considered and is associated with total yield-changes of approximately 2\%. Uncertainties stemming from the modeling of the parton showering in simulation are obtained by varying the scale at which the parton shower is evaluated. The effect of these are found to be small, with a negligible effect on process yields. Per process, uncertainties associated with the choice of the renormalisation and factorisation scale are included as well by varying each scale by a factor 0.5 and 2, respectively. Through this, a yield changing effect associated with the factorisation (normalisation) scale uncertainty of approximately 2.5\% (5\%) is introduced for background processes. The signal contributions deviate significantly from this, with the factorisation and normalisation uncertainties producing yield changes of approximately 20\% for the signal processes. Finally, an uncertainty originating from the branching ratio $\mathcal{B}(H\rightarrow ZZ \rightarrow 4\ell)~$\cite{HiggsHandook} is considered, which corresponds to a 2\% normalisation uncertainty. Additionally, uncertainties stemming from the $K$ factors applied to the assumed ZZ production cross section are also applied. These are 1.5\% and 4\% for gluon and quark-induced ZZ production, respectively.   \cite{qqZZkFactor,ggZZkFactor}.

\subsection{Expected constraints}
\label{sec:uncertainties}
\begin{figure}
    \centering
    \includegraphics[width=\linewidth]{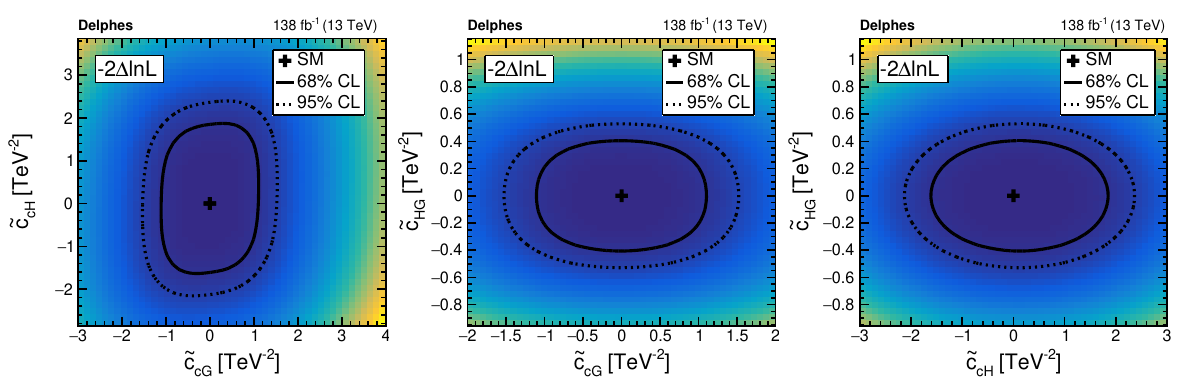}
    \caption{Results of the likelihood ($L$) scans for the simultaneous measurement of the $c_{cG}$, $c_{cH}$, and $c_{HG}$ coefficients. The contour lines show the 68\% and 95\% CL intervals associated with $-2\Delta\ln{L} = 2.3$ and $-2\Delta\ln{L} = 5.9$ respectively.}
    \label{fig:2DScans}
\end{figure}

Using the described method, expected 95\% CL upper limits on $\tilde{c}_{cG}$ and $\tilde{c}_{cH}$ are set for a scenario in which the full Run-2 dataset of the CMS detector is used. The results are summarized in Table~\ref{table:results}. The constraint on $\tilde{c}_{cG}$ is evaluated with an additional criterion on the mass of the Higgs+jet system $m_\mathrm{H+\mathrm{jet}}<1$~TeV, to ensure that the perturbativity requirement discussed in Section~\ref{sec:validity} is satisfied. Additionally, a statistical extrapolation to the anticipated size of the HL-LHC dataset is performed. These results are also shown in Table~\ref{table:results}. It should be noted that this extrapolation does not account for significant changes in the CMS detector and LHC operating conditions during the HL-LHC era \cite{HLLHC-CMS}.\\
\begin{table}[h!]
 \centering
  \caption{The expected 95\% CL upper limits derived for a LHC Run-2 and HL-LHC scenario for the $\tilde{c}_{cG}$ and $\tilde{c}_{cH}$ Wilson coefficients.}
  \label{table:limits}
  \begin{tabular}{llccc}
    \toprule
    Wilson & Expected upper limit & Expected upper limit\\
    coefficient & at 138 fb$^{-1}$ [TeV$^{-2}$] & at 3000 fb$^{-1}$ [TeV$^{-2}$]\\
    \midrule
    $\tilde{c}_{cG}$ & 1.36 & 0.56 \\
    $\tilde{c}_{cH}$ & 2.31 & 1.76 \\
    \bottomrule
  \end{tabular}
  \label{table:results}
\end{table}
\\
The limits obtained for the CMD operator $\hat{O}_{cG}$ with $M_{cut} < 1$ TeV are consistent with the validity requirements displayed in Fig.~\ref{fig:validityCMD} for both of the scenarios given in Table~\ref{table:results}. In fact, by applying Eq.~\eqref{eq:perturbativityCMD}, the estimated limits respect the perturbativity requirement for $\epsilon > 0.09$.  \\
\\
To make a comparison to the previously discussed limits on $\tilde{c}_{HG}$, originating from an analysis of Higgs boson production and decay rates as reported in \cite{CMSHiggsCombinedMeasurements}, the sensitivity of this analysis to $\tilde{c}_{HG}$ is also estimated. Here, a 95\% CL upper limit of $\tilde{c}_{HG} = 0.52$ TeV$^{-2}$ is derived for the full Run-2 dataset of the CMS detector. This indicates the analysis presented in this work is about two orders of magnitude less sensitive to $\tilde{c}_{HG}$ than the analysis presented in \cite{CMSHiggsCombinedMeasurements}, making the latter much more suitable for constraining $\tilde{c}_{HG}$.\\
\\
We have also considered SMEFT scenarios where two operators $\hat{O}_1$ and $\hat{O}_2$ are allowed to have non-zero Wilson coefficients at the same time, for the LHC Run-2 scenario. The cH process now receives contributions from the interference and quadratic terms of both operators, in addition to a term corresponding to the interference between the two operators $\propto c_1 c_2$. To estimate the sensitivity to both coefficients simultaneously, a likelihood-based scan on a grid of $(c_1, c_2)$ values was performed using the leading jet $p_{T}$ as the observable of interest. From this grid, contours are obtained associated with the 68\% and 95\% confidence intervals. The results are shown in Fig.~\ref{fig:2DScans}.\\
\\
In the simultaneous measurement of $\tilde{c}_{cG}$ and $\tilde{c}_{cH}$, it is observed that the expected upper limits on $\tilde{c}_{cG}$ remain unchanged across most of the range in which $\tilde{c}_{cH}$ is varied, which suggests that the associated limits are largely uncorrelated. This is expected, as we observe that the shape effects introduced by $\tilde{c}_{cG}$ on the leading jet $p_{T}$ spectrum are generally distinguishable from the normalization effect introduced by $\tilde{c}_{cH}$, as seen in Fig.~\ref{fig:LeadingJetPt}. On the other hand, when $\tilde{c}_{HG}$ is allowed to have non-zero values together with $\tilde{c}_{cG}$ or $\tilde{c}_{cH}$, we observe that the 95\% and 65\% CL contour lines become more elliptical, indicating some correlation between the expected upper limits of the Wilson coefficients. This is understood to be the result of the contributions of $\tilde{c}_{HG}$ to the leading jet $p_{T}$ bins in Fig.~\ref{fig:LeadingJetPt} overlapping with those of $\tilde{c}_{cG}$ or $\tilde{c}_{cH}$. This suggests that $\tilde{c}_{HG}$ introduces a degeneracy in the simultaneous measurement with $\tilde{c}_{cG}$ or $\tilde{c}_{cH}$. The breaking of this degeneracy would likely require taking advantage of subtle differences between the operators across multiple observables, which indicates that the use of multivariate techniques is preferred. This is, however, outside the scope of this work, and is left to future continuations.
\\
\\
The leading uncertainties impacting the sensitivity achieved for the derived constraints are estimated using the `Impacts' feature of the \textsf{Combine} tool. These are predominantly theoretical in nature. Especially the uncertainty associated with the PDF set as well as the uncertainties on the renormalisation and factorisation scale of the signals and the ggH process play a significant role. Leading experimental uncertainties include the uncertainties associated with the jet energy scale and jet energy resolution calibration. \\
\\
The use of cut-based jet flavor tagging criteria, using a parametrisation of the \textsf{DeepJet} \cite{DeepJet} algorithm, was tested to gauge their impact on the upper limits that are derived. However, no gain in analysis sensitivity was observed, likely due to the limited efficiencies with which c jets may be identified. This, however, may change in an experimental context where the full jet flavor discriminator spectra are available. Given this, and the ongoing improvements in jet flavor tagging efficiency, such as those achieved with the \textsf{ParticleTransformer} algorithm \cite{ParticleTransformer}, we recommend verifying the impact of jet flavor tagging in an experimental context. \\
\\
Other analysis improvements may come from the inclusion of additional Higgs boson decays. The simplest extension of the presented analysis would be the inclusion of final states with electrons. However, other Higgs boson decay modes such as $H \rightarrow \gamma \gamma$ or $H \rightarrow WW^*$ may also be explored, though the background processes associated with these decay modes change significantly. Additional potential improvements, not explored in this work, also stem from the use of multivariate analysis to combine the information of multiple discriminating observables and as such better separate EFT signals from the SM background.

\section{Summary and outlook}
\label{sec:summary}
In summary, this work discusses the proton-proton collision process in which a Higgs boson is produced in association with a charm quark (cH), and the unique, and yet unexplored way, in which it may be used to probe BSM effects using an EFT approach. A number of EFT operators that affect the cH process are discussed, including the $\hat{O}_{cG}$, $\hat{O}_{cH}$, and $\hat{O}_{HG}$ operators. Their relevance to this work is considered and their contributions to the cH cross section is estimated. Additionally, perturbativity bounds for the relevant operators are derived in the context of a minimal UV-complete extension of the SM, to ensure that any derived constraints on Wilson coefficients do not spoil the perturbativity of this EFT interpretation. Subsequently, an event reconstruction to identify cH event candidates is presented, and observables are identified that are well-suited as discriminators in the search for deviations from the SM associated with non-zero values of the Wilson coefficients $\tilde{c}_{cG}$ and $\tilde{c}_{cH}$. Finally, the statistical evaluation that is applied to cH candidate events is discussed along with a set of systematic uncertainty sources that are expected to be important in the measurement. The results of this evaluation is presented, with expected 95\% CL upper limits of $\tilde{c}_{cG} = 1.26$ TeV$^{-2}$ and $\tilde{c}_{cH}=2.49$ TeV$^{-2}$ obtained for a scenario in which the full dataset of events recorded by the CMS detector during the LHC Run-2 data-taking period is used. Additionally, expected 95\% CL upper limits of $\tilde{c}_{cG} = 0.52$ TeV$^{-2}$ and $\tilde{c}_{cH}=1.88$ TeV$^{-2}$ are set in an extrapolation to a High-Luminosity LHC scenario. These expected upper limits in single-operator scenarios are complemented by a scenario that considers the combined effect of pairings of two operators. The sensitivity of the analysis in this scenario is estimated using  95\% confidence interval contours in a likelihood-based scan of Wilson coefficient values. Finally, the leading uncertainty sources impacting the sensitivity of these results are identified and potential analysis improvements are briefly discussed. \\
\\
Looking towards Run-3 of the LHC, as well as ultimately the HL-LHC, the presented analysis strategy may offer interesting insight into constraining potential new physics effects that may be uniquely probed in cH events. We thus recommend the presented ideas be investigated and further developed in an experimental context.

\acknowledgments
This work has been supported by the FWO Research Foundation (project G0C3823N) and the Interuniversity BOF Projects program (project iBOF/23/074) in Flanders. 

\bibliographystyle{JHEP}
\bibliography{biblio.bib}

\end{document}